# Properties of Turbulence with Zero and Nonzero Helicity

E. Golbraikh

Physics Department of Ben-Gurion University of the Negev, POB 653

Beer-Sheva 84105, Israel

Tel. +972-8-6428228/Fax: +972-8-6280467/E-mail: golbref@bgu.ac.il

## Abstract

The present paper deals with the study of spectral properties of the helical mode of uniform isotropic turbulence in the presence and in the absence of mean helicity. It is shown that even in the absence of mean helicity, the helicity of individual helical modes affects the spectral behavior of turbulence. Besides, it is shown that both with and without mean helicity, intermediate characteristic scales connected with the presence of helicity in helical modes exist. The existence of such scales makes it possible to solve the problem of helicity dissipation divergence for individual helical modes.

*PACS:* 67.40.Vs, 67.55.Fa, 94.05.Lk

*Keywords:* Turbulence, Helical turbulence, Helicity dissipation

## Introduction

Helicity as an integral part of turbulent flows was first considered about half a century ago [1, 2]. However its role in turbulent flow behavior remains, in a certain sense, mysterious and not clear enough (see, for example, [3-5] and references therein). Being, side by side with energy, an integral of motion of the Euler equation, helicity plays an essential role in the study of various processes of large-scale structures generation and behavior both in hydrodynamic and magnetohydrodynamic flows [5-7]. At the same time, today it is clear that turbulent cascade is typical not only of energy, but also of helicity [3, 9-11]. On the other hand, the interference of energy and helicity transfer affects turbulent correlation characteristics in different scales [11, 12]. Thus, helicity is an essential element in the study of turbulence evolution.

A two-point velocity correlator of a homogeneous and isotropic incompressible turbulent flow with violated mirror symmetry can be represented in a general form (see, e.g., [13]):

$$\langle u_i(\mathbf{r},t) u_j(0,t) \rangle = A(r)\delta_{ij} + B(r)r_i r_j + C(r)\varepsilon_{ijk} r_k \tag{1}$$

where $u_i(\mathbf{r},t)$ is a fluctuating component of the velocity field in the point $\mathbf{r}$ at the moment $t$, $A(r)$, $B(r)$ and $C(r)$ are functions depending on the modulus of $\mathbf{r}$, $\varepsilon_{ijk}$ is a fully antisymmetric unit tensor.

Fourier-presentation of this correlator in a general form is:

$$\langle u^*_i(k,t) u_j(k,t) \rangle = \frac{E(k,t)}{4\pi k^2}\left(\delta_{ij} - \frac{k_i k_j}{k^2}\right) + i\varepsilon_{ikj}\frac{H(k,t)}{8\pi k^4} k_k \tag{2}$$

where $E(k,t)$ and $H(k,t)$ are energy and helicity densities, respectively, and $k$ is the wave vector. In this case, both energy $\varepsilon$ and helicity $\eta$ fluxes exist in the system. If they are governing parameters of turbulence together with kinematic viscosity, it leads to the formation of another scale $l_\eta \sim \frac{\varepsilon}{\eta}$, besides the characteristic Kolmogorov dissipation scale $l_\nu$ [8]. In compliance with [12], $l_\eta$ can be defined as a scale separating the influence of external and dissipative scales on the behavior of the energy (helicity) correlator in the inertial interval. On the other hand, as we know, helicity is a measure of mirror symmetry disturbance in a turbulent flow. However, "latent asymmetry" of the flow could affect even the mirror symmetry turbulence. In the present work we study some helical properties of the mirror symmetry turbulence and their connection with the non-zero helicity case. At the same time, we show that besides $l_\eta$, a new characteristic scale $l_B = \sqrt{l_\nu l_\eta}$ arises, which determines the onset of viscosity influence.

**Helical Modes**

If the mirror symmetry of a flow is not violated, then $H(k,t) = 0$. However, it does not mean that helical vortices do not form in the flow. Just as previously, we can expand the velocity and vortex fields in helical waves $h_s(\mathbf{k})\exp(i\mathbf{kr})$ (see, for example, [3, 13, 14]), i.e.

$$u(\mathbf{r},t) = \sum_\mathbf{k} \sum_s a_s(\mathbf{k},t) h_s(\mathbf{k}) \exp(i\mathbf{kr}) \text{ and}$$

$$\omega(\mathbf{r},t) = rot(u(\mathbf{r},t)) = \sum_\mathbf{k} \sum_s s|\mathbf{k}| a_s(\mathbf{k},t) h_s(\mathbf{k}) \exp(i\mathbf{kr}).$$

In this case, spectral densities of energy and helicity can be represented as a sum and difference of their positively defined helical components $E^\pm(k,t)$ and $H^\pm(k,t)$, which are defined as follows (we have used normalization in compliance with [3]):

$$E^\pm(k,t) = \frac{1}{2} \sum_\mathbf{p} \langle |a_\pm(\mathbf{p},t)|^2 \rangle \delta(k - |p|)$$

and

$$H^\pm(k,t) = \sum_\mathbf{p} |\mathbf{p}| \langle |a_\pm(\mathbf{p},t)|^2 \rangle \delta(k - |p|)$$

and interconnected as follows: $H^\pm(k,t) = 2k E^\pm(k,t)$, whereas

$$E(k,t) = E^+(k,t) - E^-(k,t); \quad H(k,t) = H^+(k,t) - H^-(k,t). \tag{3}$$

where $\langle ... \rangle$ above denotes averaging over an ensemble.

If $H(k,t) = 0$ and $H^+(k,t) = H^-(k,t)$, then $E^+(k,t) = E^-(k,t) = \frac{1}{2}E(k,t)$.

Balance equations for spectral densities of energy and helicity lead to the following equations for $E^{\pm}(k,t)$ and $H^{\pm}(k,t)$ [3]:

$$\frac{\partial E^{\pm}}{\partial t} = T_E^{\pm}(k,t) - 2vk^2 E^{\pm} + F_E^{\pm}; \qquad \frac{\partial H^{\pm}}{\partial t} = T_H^{\pm}(k,t) - 2vk^2 H^{\pm} + F_H^{\pm} \qquad (4)$$

where $T_X^{\pm}(k,t)$ is a flux of the respective component over the spectrum, and $F_X^{\pm}(k,t)$ is a function of its source. It should be noted that equalities $T_E^{\pm}(k,t) = \frac{1}{2}\left[T_E \pm \frac{T_H}{2k}\right]$ and $T_H^{\pm}(k,t) = \frac{1}{2}[2kT_E \pm T_H]$ (at $H(k,t) \neq 0$) [3], whence it follows that $T_H^{\pm} = 2kT_E^{\pm}$ (and $F_H^{\pm} = 2kF_E^{\pm}$), lead to the system (4) degeneration, which is *wrong* in a general case in all the intervals. In the case of a steady state and $H(k,t) = 0$, in the *inertial* interval $T_E^{\pm}(k,t) = \frac{1}{2}T_E$ and $T_H^{\pm}(k,t) = kT_E$.

If $H(k,t) = 0$ and assuming that helicity generation and dissipation are sufficiently spaced in scales, we obtain from the second equation (4) that $T_H^{+}(k,t) = T_H^{-}(k,t)$ within the dissipative range, and $T_H^{+}(k,t) - T_H^{-}(k,t) = -(F_H^{+}(k,t) - F_H^{-}(k,t))$ in the range of larger scales. Assuming that $F_H^{+}(k,t) = F_H^{-}(k,t)$ (since there are no reasons for asymmetry), we obtain that $T_H^{+}(k,t) = T_H^{-}(r,t)$ in all scales. Similarly, we obtain the same for $T_E^{+}(k,t) = T_E^{-}(r,t)$. Note that $T_H^{\pm} \neq 2kT_E^{\pm}$ in all scales simultaneously, in contrast to [3]. Hence, the flow is helical with a maximal helicity of each helical component. Here, however, the helicity is connected only with the flow twist in helical vortices, and not with their knots that generate nonzero mean helicity in flows with violated mirror symmetry [2,15].

Integrating equations (4) over $k$ in case of a steady state and $H(k,t) = 0$, we can obtain that

$$D_E^\pm = F_E^\pm = \frac{\varepsilon}{2} \text{ and } D_H^\pm = F_H^\pm = \eta^\pm, \text{ where } D_X^\pm = 2\nu \int_0^\infty k^2 X^\pm(k,t)dk \text{ and}$$

$$F_X^\pm = \int_0^\infty F_X^\pm(k,t)dk \text{ (where } X = E \text{ or } H\text{). In case of violated symmetry, it is just}$$

$\eta_H = \eta^+ - \eta^-$ that is conserved in the total flow in a steady state, because in this case

$\int_0^\infty T_H^\pm(k,t)dk \neq 0$, whereas $\int_0^\infty T_H(k,t)dk = 0$. At a mirror symmetry of the flow, $\eta_H = 0$ and the flows are conserved in each mode.

Reverting to Eq. (2), we introduce, according to [13],

$$\langle u^*_i(k,t)u_j(k,t)\rangle^+ = \frac{E^+(k,t)}{4\pi k^2}\left(\delta_{ij} - \frac{k_i k_j}{k^2}\right) + i\varepsilon_{ikj}\frac{H^+(k,t)}{8\pi k^4}k_k$$
$$\langle u^*_i(k,t)u_j(k,t)\rangle^- = \frac{E^-(k,t)}{4\pi k^2}\left(\delta_{ij} - \frac{k_i k_j}{k^2}\right) - i\varepsilon_{ikj}\frac{H^-(k,t)}{8\pi k^4}k_k$$

(8)

whose sum gives Eq. (2), while their difference is

$$\langle u^*_i(k,t)u_j(k,t)\rangle^+ - \langle u^*_i(k,t)u_j(k,t)\rangle^- = \frac{H(k,t)}{8\pi k^3}\left(\delta_{ij} - \frac{k_i k_j}{k^2}\right) + i\varepsilon_{ikj}\frac{E(k,t)}{4\pi k^3}k_k \qquad (9)$$

i.e., helicity is a measure of energy difference in heteropolarized helical modes.

At $H(k,t) = 0$ we obtain

$$\langle u^*_i(k,t)u_j(k,t)\rangle^+ - \langle u^*_i(k,t)u_j(k,t)\rangle^- = i\varepsilon_{ikj}\frac{E(k,t)}{4\pi k^3}k_k$$

$$\langle u^*_i(k,t)u_j(k,t)\rangle^+ + \langle u^*_i(k,t)u_j(k,t)\rangle^- = \frac{E(k,t)}{8\pi k^2}\left(\delta_{ij} - \frac{k_i k_j}{k^2}\right)$$

(10)

and

$$\langle u^*_i(k,t)u_j(k,t)\rangle^+ = \frac{E(k,t)}{16\pi k^2}\left(\delta_{ij} - \frac{k_i k_j}{k^2} + 2i\varepsilon_{ikj}\frac{k_k}{k}\right)$$

$$\langle u^*_i(k,t)u_j(k,t)\rangle^- = \frac{E(k,t)}{16\pi k^2}\left(\delta_{ij} - \frac{k_i k_j}{k^2} - 2i\varepsilon_{ikj}\frac{k_k}{k}\right)$$

(11)

In this case only, according to Eq. (11), velocity correlators pass into each other at a mirror reflection ($i \leftrightarrow j$). Thus, for every helical mode, the flow is helical, and energy (and helicity) transfer therein occurs *without* mixing. Here the helicities are maximal in each component $H^\pm(k,t) = 2kE^\pm(k,t)$, and the total flow is reflection symmetric. In the present case, energy becomes a macroparameter of the flow, and helicity is a latent parameter. (However, it is a perfectly real parameter for each of helical modes).

**Compensated and non-compensated modes**

Note that even the form of Eqs.(1), (2) and (9) of isotropic turbulence correlators points to the possibility of considering a flow in the absence of symmetry as a set of reflection symmetric and asymmetric parts or, in other words, helical compensated and non-compensated parts. The compensated part of helicity corresponds to $H(k) = 0$ and is analogous to the idea of weakly interacting vortons [4], while the non-compensated part is connected with helical structures having nonzero knots. This part describes the interaction between various helical components enhancing, in certain cases, one of them.

Thus, $2k(E^+(k,t) - E^-(k,t)) = H(k)$ is a non-compensated part of helical modes with the energy density (see also Eq. (9))

$$E_H(k,t) = |E^+(k,t) - E^-(k,t)| = \left|\frac{H(k,t)}{2k}\right|. \tag{12}$$

Since for the compensated part $H(k,t) = 0$, energy density of each of its parts is

$$E_f(k,t) = \frac{1}{2}(E^+(k,t) + E^-(k,t) - E_H(k,t)) = \frac{1}{2}(E(k) - E_H(k,t)) \tag{13}$$

*Asymptotic model relations*

Assuming that in the inertial interval [10, 18, 19]

$$E(k) = C_E \left(\frac{\varepsilon^2}{\eta}\right)^{2/3} \left(\frac{\eta}{\varepsilon}\right)^{\delta_E} k^{-(\delta_E+1)} \text{ and } H(k) = C_H (\eta\varepsilon)^{1/3} \left(\frac{\eta}{\varepsilon}\right)^{\delta_H} k^{-(\delta_H+1)}$$

(here $\delta_E = \frac{4+6\alpha_1}{3(1-\alpha_1)}$ and $\delta_H = \frac{1-6\alpha_1}{3(1-\alpha_1)}$; the parameter $-\frac{2}{3} < \alpha_1 < \frac{1}{6}$), then, since

$$E^\pm(k) = \frac{1}{2}(E(k) \pm \frac{H(k)}{2k}) \text{ and } H^\pm(k) = k(E(k) \pm \frac{H(k)}{2k}),$$

we obtain

$$E^\pm(k) = \frac{1}{2} C_E \left(\frac{\varepsilon^2}{\eta}\right)^{2/3} \left(\frac{\eta}{\varepsilon}\right)^{\delta_E} k^{-(\delta_E+1)} \left(1 \pm \frac{C_H}{2C_E} \left(\frac{\eta}{\varepsilon}\right)^{\delta_H-\delta_E+1} k^{-(\delta_H-\delta_E+1)}\right) \tag{14a}$$

$$H^{\pm}(k) = \frac{1}{2}C_E\left(\frac{\varepsilon^2}{\eta}\right)^{2/3}\left(\frac{\eta}{\varepsilon}\right)^{\delta_E} k^{-\delta_E}\left(1 \pm \frac{C_H}{2C_E}\left(\frac{\eta}{\varepsilon}\right)^{\delta_H-\delta_E+1} k^{-(\delta_H-\delta_E+1)}\right) \quad (14b)$$

which exactly coincide with expressions derived in [3, 13, 16] at $\alpha_1 = -\frac{1}{4}$ (Kolmogorov's case).

On the other hand, according to Eqs. (12) and (13),

$$E_H(k,t) = \frac{C_H}{2}(\eta\varepsilon)^{1/3}\left(\frac{\eta}{\varepsilon}\right)^{\delta_H} k^{-(\delta_H+2)} \quad (15a)$$

and

$$E_f(k,t) = \frac{1}{2}C_E\left(\frac{\varepsilon^2}{\eta}\right)^{2/3}\left(\frac{\eta}{\varepsilon}\right)^{\delta_E} k^{-(\delta_E+1)}\left(1 - \frac{C_H}{2C_E}\left(\frac{\eta}{\varepsilon k}\right)^{\delta_H-\delta_E+1}\right) \quad (15b)$$

It can be easily shown that in Kolmogorov's case only, when $\delta_E = \delta_H = 2/3$,

$$E_f(k,t) = \frac{1}{2}C_E\varepsilon^{2/3}k^{-5/3}\left(1 - \frac{C_H}{2C_E}\frac{\eta}{\varepsilon}k^{-1}\right) \text{ and } E_H(k,t) = \frac{C_H}{2}\varepsilon^{-1/3}\eta^{2/3}k^{-8/3} \quad (16)$$

In the helical case, when $\alpha_1 = 0$, $\delta_E = \frac{4}{3}$ and $\delta_H = \frac{1}{3}$, we obtain:

$$E_f(k,t) = \frac{1}{2}C_E\eta^{2/3}k^{-7/3}\left(1 - \frac{C_H}{2C_E}\right) \text{ and } E_H(k,t) = \frac{C_H}{2}\eta^{2/3}\left(\frac{\eta}{\varepsilon}\right)^{1/3} k^{-7/3} \quad (17)$$

i.e. the background turbulence has a helical spectrum [5, 8], just as $E_H(k,t)$ and $E(k,t)$.

Thus, the total helicity takes part in the energy transfer as an additional channel, and for each $k$, energy emission or absorption by components of a symmetric field takes place.

Meanwhile, helicity growth in certain scales (see, e.g., [5, 19, 20]) reflects the increasing energy disbalance among helical components and enhancement of one of them. Since this disbalance decreases with growing $k$ according to Eq. (12) (see also [3, 24]), generation of helical vortices in a preferential direction occurs in large scales.

**Dissipation**

In case of reflection-symmetric turbulence, the main parameter of the system is the energy flux $\varepsilon$ (Kolmogorov's case) only; besides, $\eta \to 0$ and $\delta_E = 2/3$. Then, according to Eq. (14), $E^\pm(k) \sim \varepsilon^{2/3} k^{-5/3}$ and $H^\pm(k,t) \sim \varepsilon^{2/3} k^{-2/3}$ for individual helical components of the flow. Then, as follows from [3,16], a problem of integral dissipation divergence arises

$$\widetilde{D}_H^\pm \approx 2\nu \int_0^{k_\nu} k^3 E^\pm(k) dk \text{ (where } k_\nu \text{ corresponds to an inverse Kolmogorov's scale } l_\nu = \left(\frac{\nu^3}{\varepsilon}\right)^{1/4} \text{ )}$$

at $\nu \to 0$, if -5/3 spectrum is used *for each of helical components* only.

This problem of divergence is due to the fact that we are unaware of the form of the function $E(k)$ (and $E^\pm(k)$) in the *near dissipation region* and make use of the function $E(k)$ from the inertial interval. Many attempts have been made to establish the behavior of $E(k)$ in this region (see [21, 22] and references therein). As follows from these works, $E(k)$ spectrum becomes sufficiently steep, which eliminates divergences. Here the form of $E(k)$ is not purely exponential (for example, it can be of $E \sim k^{-\delta_E}/(a+(k/k_\nu)^\gamma)^\lambda$ type, where $a$, $\gamma$ and $\lambda$ are constants, $\gamma$ and $\lambda$ being greater than zero, and $\delta_E$ is the spectral exponent in the inertial interval), which acquires the power function form $E \sim k^{-\delta_E}$ only in the limiting case ($k \ll k_\nu$).

This is confirmed by experimental data. For example, it is evident from a well-known plot of the dependence of longitudinal velocity increments spectrum on the wave number normalized to $k_v$ presented in [23] (as well as in [21, 22]) that at $k \sim k_v$ the slope of the curve is much steeper that in the inertial region, where it is close to Kolmogorov's law of 5/3. As follows from experimental data, the scale of the transition from the inertial range spectrum $l_B$ to the dissipative range spectrum corresponds to $\sim (10 \div 60) l_v$.

It is noteworthy (see [22] and references therein) that a multifractal approach also leads to a steeper power spectrum in the near dissipation region. Therefore, we assume that

$$\widetilde{D}_H^\pm \sim \nu \left( \int_0^{k_B} E(k) k^3 dk + \int_{k_B}^{k_v} E(k) k^3 dk \right) \tag{18}$$

where $k_B$ corresponds to the scale where the transition from one spectral slope to another takes place (a bend appears).

Approximate formula of spectral energy transfer based on various assumptions concerning the relations between energy transfer $R_E(k,t) = \int_0^k T_E(k',t)dk'$ and spectral energy density $E(k)$ are described in detail in [21]. It follows from various hypotheses about the form of the function $R(k, E(k), t)$ that at $k \ll k_v$ $E(k) \to E(k) \sim k^{-5/3}$, and at $k \sim k_v$

$E(k) \to E(k) \sim k^{-5/3} \varphi\left(\dfrac{k}{k_v}\right)$. Here $\varphi(x)$ is an unknown function of $x$. As noted above, if helicity transfer, side by side with energy transfer over the spectrum, plays an essential role, then, besides the characteristic dissipation scale $l_\nu$, another scale $l_\eta \sim \dfrac{\varepsilon}{\eta^\pm}$ appears. Then it is necessary to generalize the function $\varphi(x)$ to this case. Here the asymptotic relation of the spectral density (or velocity correlator) between the dissipative and inertial intervals should be retained. In this case,

$$\varphi\left(\frac{k}{k_\nu}\right) \to \varphi\left(\frac{k}{k_\nu}, \frac{k}{k_\eta}\right) \tag{19}$$

As a result of limiting transitions, $\varphi(x)$ becomes $\varphi\left(\dfrac{k}{k_\eta}\right)$ at $\dfrac{k}{k_\nu} \ll \dfrac{k_\eta}{k}$ and $\varphi\left(\dfrac{k}{k_\nu}\right)$ at $\dfrac{k}{k_\nu} \gg \dfrac{k_\eta}{k}$ in the inertial and dissipative intervals, respectively. Therefore, the function $\varphi\left(\dfrac{k}{k_\nu}, \dfrac{k}{k_\eta}\right) \to \varphi\left(\dfrac{k}{k_\nu}, \dfrac{k_\eta}{k}\right)$ and the transition scale $l_B \sim \dfrac{1}{k_B}$ can be determined from the equality $\dfrac{k}{k_\nu} = \dfrac{k_\eta}{k}$

$$k_B = \sqrt{k_\eta k_\nu} = \left(\eta^\pm\right)^{\frac{1}{2}} (\varepsilon\nu)^{-\frac{3}{8}} \tag{20}$$

Unfortunately, experimental data on helicity in turbulent flows is very scarce. Therefore, the main index of helicity influence on a turbulent spectrum is the presence of -7/3 exponent in the inertial range [5, 8]. On the basis of experimental data on turbulent velocity fluctuations measured in a mercury flow in a transverse magnetic field (MHD flow) obtained in [25] and in the atmospheric boundary layer (BL) [26], we can evaluate $k_B$. It follows from the experiment that energy and helicity fluxes amounted to $\varepsilon_{MHD} \approx (7 \div 10)10^{-6} \, m^2/s^3$, $\varepsilon_{BL} \approx 0.28 \, m^2/s^3$, $\eta_{MHD} \approx (2 \div 4) \, 10^{-4} \, m/s^3$ and $\eta_{BL} \approx 0.28 \, m/s^3$, which corresponds to the scales $l_\eta^{MHD} \approx 0.03 \, m$, $l_\eta^{BL} \approx 1 \, m$, $l_\nu^{MHD} \approx 1 \, 10^{-4} \, m$, $l_\nu^{BL} \approx 7.3 \, 10^{-4} \, m$, $l_B^{MHD} \approx 1.8 \, 10^{-3} \, m$ and $l_B^{BL} \approx 2.7 \, 10^{-2} \, m$ ($\nu_{mercury} \approx 1.15 \, 10^{-7} \, m^2/s$ and $\nu_{air} \approx 2 \, 10^{-6} \, m^2/s$). Thus, $\dfrac{l_B^{MHD}}{l_\nu^{MHD}} \approx 18$ and $\dfrac{l_B^{BL}}{l_\nu^{BL}} \approx 37$, which well agrees with experimental data given in [23], as well as in [21, 22]. It is noteworthy that in Ref. [25], results of different atmospheric experimental data processing are also presented. However, these data refer to an external scale range, which makes it impossible to use them for evaluating $l_B$. Another laboratory experiment allowing, in our opinion, the evaluation of $l_B$, is described in Ref. [27] dealing with water rotation. In this experiment, $l_\eta \approx 0.07 \, m$, but energy flux along the spectrum was not determined. On the other hand, if the characteristic value $l_\nu \approx \dfrac{\overline{V}}{\nu_{water}} \approx 0.0001 \, m$ is used ($\overline{V} \approx 0.01 \, m/s$ being characteristic velocity, and $\nu_{water} \approx 10^{-6} \, m^2/s$), then $l_B \approx 2.6 \, 10^{-3} \, m$, leading to $\dfrac{l_B}{l_\nu} \approx 26$, which also agrees with data of Ref. [23].

Thus, it is necessary to substitute the upper integration limit in Eq. (18) with $k_B$, which leads to

$$D_H^\pm \sim \nu k_B^{7/3} \propto \nu^{1/8} \xrightarrow[\nu \to 0]{} 0 \qquad (21)$$

Hence, in Kolmogorov's case, the first integral (12) does not diverge with growing Reynolds number.

To assess the second integral in (18), we examine the following. According to

$$D_H^\pm = \eta^\pm = 2\nu \int_0^\infty k^2 H^\pm(k,t)dk,$$

$$2\nu \int_{k_B}^\infty k^2 H^\pm(k,t)dk = \eta^\pm - 2\nu \int_0^{k_B} k^2 H^\pm(k,t)dk = \eta^\pm - A(\eta^\pm)^{7/6} \varepsilon^{-5/24} \nu^{1/8} \qquad (22)$$

where $A = \dfrac{6}{7} C_E^\pm$. We assume that the function $k^2 H(k)$ has a maximum in the vicinity of $k = k_\nu$, and the dissipation is mainly concentrated in the interval $k_B \div k_\nu + k_B$, which is symmetrical with respect to $k_\nu$, i.e. $2\nu \int_{k_B}^\infty k^2 H(k)dk \approx 2\nu \int_{k_B}^{k_\nu + k_B} k^2 H(k)dk \approx 4\nu \int_{k_B}^{k_\nu} k^2 H(k)dk$. Now we can estimate $D_H^\pm$:

$$\widetilde{D}_H^\pm \approx 2\nu \int_0^{k_\nu} k^3 E^\pm(k)dk \cong \frac{1}{2}\left(\eta^\pm + A(\eta^\pm)^{7/6} \varepsilon^{-5/24} \nu^{1/8}\right) \qquad (23)$$

Thus, the integral helicity dissipation does not diverge at $\nu \to 0$, although there is a bend in the helicity spectrum in the scales of the transition from the inertial to the dissipative interval connected with the characteristic scale of the inertial region $l_\eta$.

In case of violated mirror symmetry, the above reasoning cannot refer to each of helical components separately; they refer to the total energy and helicity (i.e. it is necessary to substitute all $X^{\pm}(k)$ with $X(k)$). In this case, $E^{\pm}(k) = \frac{1}{2}(E(k) \pm \frac{H(k)}{2k})$ and $H^{\pm}(k) = k(E(k) \pm \frac{H(k)}{2k})$, and the divergence $\widetilde{D}_H^{\pm} \approx 2\nu \int_0^{k_\nu} k^3 E(k) dk$ is connected with the extension of $E(k)$ dependence from the inertial interval into the pre-dissipative one. On the other hand, $l_\nu \sim \varepsilon^{\alpha_1} \eta^{-1/5(1+4\alpha_1)} \nu^{3/5(1-\alpha_1)}$ within the bounds of the asymptotic model, and the transition scale is

$$l_B = \sqrt{l_\eta l_\nu} = \left(\varepsilon^{\alpha_1+1} \eta^{-1/5(6+4\alpha_1)} \nu^{3/5(1-\alpha_1)}\right)^{\frac{1}{2}} \tag{24}$$

**Discussion**

The appearance of a characteristic transition scale in the presence of non-zero helicity of helical modes can also appear in the structure of the total energy spectrum $E(k)$. In fact, as follows from [10], we can write the spectral energy density in this interval as $E^{\pm}(k) \sim (\varepsilon \nu)^{1/2} l_\nu^{\delta_1} k^{-(\delta_1+1)}$ (since at a totally zero helicity the spectrum in the inertial interval is degenerated into Kolmogorov's spectrum, i.e. $\alpha_1 = -\frac{1}{4}$), which leads to the following dependence of the integral helicity dissipation in the interval $k_B \leq k \leq k_\nu$:

$$\widetilde{D}_H^{\pm} \sim (\nu^3 \varepsilon)^{1/2} l_\nu^{\delta_1} \frac{1}{3-\delta_1} k^{2-\delta_1} \Big|_{k_B}^{k_\nu} \tag{25}$$

(where $\delta_1$ is the spectral index in this region in $r$-space). In the lower limit, we obtain $\delta_1 \geq 2/3$ from the condition of dissipation finiteness with growing Reynolds number. This points to the fact that energy spectrum in this region should be at least the same as in the inertial interval or even steeper. On the other hand, since $\widetilde{D}_H^{\pm} > 0$, then, according to (25), $\delta_1 < 3$. Hence, for Kolmogorov's turbulence, $2/3 \leq \delta_1 < 3$.

In case of a violated mirror symmetry of the total flow, the spectral index behavior is somewhat different. As follows from [3, 18, 24], the divergence of the total helicity dissipation with growing Reynolds number is absent and appears in helical waves only

$H^{\pm}(k) = k\left(E(k) \pm \dfrac{H(k)}{2k}\right)$ owing to the first term of this equality. According to [10], we can write $E_{diss}(k) \sim \varepsilon^{-2\alpha_1} \eta^{2/5(1+4\alpha_1)} \nu^{2/5(2+3\alpha_1)} l_\nu^{\delta_1} k^{-(\delta_1+1)}$ for the interval $k_B \leq k \leq k_\nu$, which leads to

$$\widetilde{D}_H^{\pm} \sim \nu \varepsilon^{-2\alpha_1} \eta^{2/5(1+4\alpha_1)} \nu^{2/5(2+3\alpha_1)} l_\nu^{\delta_1} \dfrac{1}{3-\delta_1} k^{2-\delta_1} \Big|_{k_B}^{k_\nu}. \tag{26}$$

In the upper limit, under the condition of $\widetilde{D}_H^{\pm}$ finiteness at $\mathrm{Re} \to \infty$, we arrive at a relationship $\alpha_1 \geq -\dfrac{1}{4}$, i.e. $-\dfrac{1}{4} \leq \alpha_1 < \dfrac{1}{6}$. In the lower limit, however, we obtain, $\delta_1 \geq \dfrac{4+6\alpha_1}{3(1-\alpha_1)}$, but in this case the boundary value of $\delta_1$ is a function of $\alpha_1$, as shown in Fig. 1. Thus, at each $\alpha_1$ value, the minimal value of $\delta_1$ corresponds to $\delta_E$ value in the inertial interval and can exceed it, i.e. $\dfrac{4+6\alpha_1}{3(1-\alpha_1)} \leq \delta_1 < 3$.

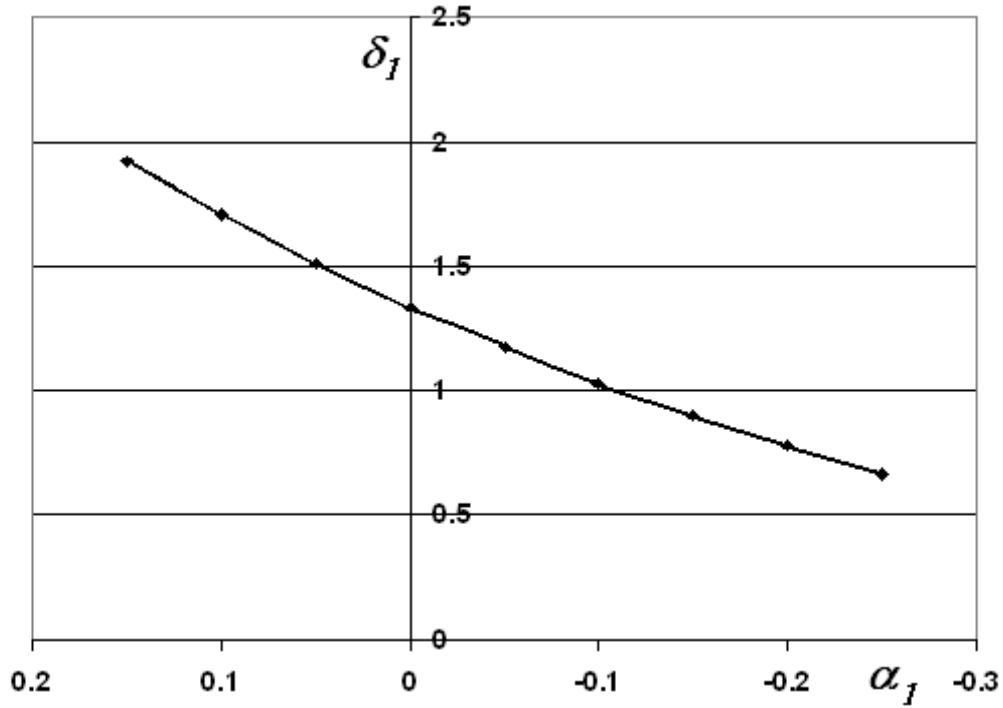

Fig. 1. Dependence of the boundary $\delta_1$ value on $\alpha_1$.

Thus, homogeneous and isotropic turbulent flows with non-violated mirror symmetry possess a latent helicity, which can affect their properties. In particular, it becomes evident in the behavior of the energy spectrum in the pre-dissipative interval of scales. In this case, an additional characteristic scale of a transition to the dissipative interval arises, where the energy density spectrum either retains its slope or becomes steeper. Besides, helicity dissipation in helical modes does not diverge with growing Reynolds number.

It is also shown that at a violated mirror symmetry, an analogous scale appears, below which the energy spectrum becomes steeper.

**Acknowledgments**

The author is grateful to Professors A. Eidelman and N. Kleeorin for useful discussions.

**References**

1. J. J. Moreau, "Constants d'un ilot tourbillonnaire en fluide parfait barotrope," C. R. Acad. Sci. Paris **252**, 2810 (1961).

2. H.K. Moffatt, "The degree of knottedness of tangled vortex lines," J. Fluid Mech. **35**, 117 (1969).

3. Q. Chen, S. Chen, and G.L. Eyink, "The joint cascade of energy and helicity in three-dimensional turbulence," Phys. Fl. **15**, 361 (2003).

4. H.K. Moffatt and A. Tsinober, "Helicity in laminar and turbulent flow." Annu. Rev. Fluid Mech. **24**, 281 (1992).

5. H. Branover, A. Eidelman, E. Golbraikh and S. Moiseev, "Turbulence and Structures" (Academic Press, NY – Boston – London, 1999).

6. F. Krause and K.-H. Radler, "Mean-Field Magnetohydrodynamics and Dynamo Theory" (Pergamon Press, Oxford, 1980).

7. S.I. Vainshtein, Ya.B. Zel'dovich, and A.A. Ruzmaykin, "Turbulent Dynamo in Astrophysics" (Nauka, Moscow, 1980).

8. A. Brissaud, U. Frisch, J. Leorat, M. Lesieur, and A. Mazure, "Helicity cascades in fully developed isotropic turbulence," Phys. Fluids **16**, 1366 (1973).

9. S.S. Moiseev and O.G. Chkhetiani, "Helical scaling in turbulence," JETP **83**, 192 (1996).

10. E. Golbraikh and S. Moiseev, "Different spectra formation in the presence of helical transfer," Phys. Lett. A **305**, 173 (2002).

11. S. Kurien, M.A.Taylor, and T. Matsumoto, "Cascade time for energy and helicity in homogeneous isotropic turbulence," Phys. Rev. E **69**, 066313-1 (2004).

12. E. Golbraikh and A. Eidelman, "On the structure of complicated turbulent spectra," Phys. A **374**, 403 (2007).

13. M. Lesieur, "Turbulence in Fluids" (Kluwer Acad. Publ., 1991).

14. F. Waleffe, "Inertial transfers in the helical decomposition," Phys. Fluids A **5**, 677 (1993).

15. H.K. Moffatt, "The energy spectrum of knots and links," Nature **347**, 367 (1990).

16. P.D. Ditlevsen and P. Giuliani, "Dissipation in Helical Turbulence," Phys. Fluids **13**, 3508 (2001).

17. E. Golbraikh, "Helical turbulent spectra in the presents of energy andhelicity fluxes," Phys. Lett. A **354**, 214 (2006).

18. O.G. Chkhetiani, and E. Golbraikh, Helicity Spectra and Dissipation, Phys. Lett. A, V.372, 5603-5604 (2008)

19. A. Belian, Chkhetiani, O., Golbraikh, E. and Moiseev, S. 1998. Helical turbulence: turbulent viscosity and instability of the second moments. Physica A, v. 258, No. 1-2, 55-68 (1998)

20. A. Belian, , O. Chkhetiani, E. Golbraikh, and S. Moiseev, "Helical turbulence: Turbulent viscosity and instability of the second moments," Physica A **258**, 55 (1998); E. Golbraikh, O. Chkhetiani, and S. Moiseev, "The role of helicity in turbulent MHD flows," JETP **87**, 95 (1998); O. Chkhetiani, S. Moiseev, and E. Golbraikh, "Helicity generation in turbulent MHD flows," ZETF **114**, 946 (1998).

21. A.S Monin,. and A.M. Yaglom, "Statistical fluid mechanics: Mechanics of turbulence" (MIT Press, Cambridge, Mass, 1975).

22 U. Frisch, "Turbulence: The Legacy of A.N. Kolmogorov" (Cambridge University Press, Cambridge, 1995).

23 C.H. Gibson, and W.H. Schwarz, "The universal equilibrium spectra of turbulent velocity and scalar fields," J. Fluid Mech. **16**, 365 (1963).

24 B . Galanti, and A. Tsinober, "Physical space properties of helicity in quasi-homogeneous forced turbulence", Physics Letters A **352** , 141–149 (2006)

25. A. Eidelman, H. Branover and S.S. Moiseev, Helical turbulence properties in the laboratory and in nature, Proc. ETC8 , "Advance in Turbulence", Barcelona, Spain, June 27-30, 61-64 (2000)

26. P. J. Mason and King, J. C., Measurements and predictions of flow and turbulence over an isolated hill of moderate slope, Quart. J. R. Met. Soc., 111, 617–640, (1985).

27. R. W. Griffits and E.J. Hopfinger, The structure of mesoscale turbulence and horizontal spreading at ocean fronts, Deep-Sea Res., **21**, 245-269, (1984)